\documentstyle[12pt]{article}
\begin{document}
\def\thebibliography#1{\section*{REFERENCES\markboth
 {REFERENCES}{REFERENCES}}\list
 {[\arabic{enumi}]}{\settowidth\labelwidth{[#1]}\leftmargin\labelwidth
 \advance\leftmargin\labelsep
 \usecounter{enumi}}
 \def\newblock{\hskip .11em plus .33em minus -.07em}
 \sloppy
 \sfcode`\.=1000\relax}
\let\endthebibliography=\endlist

\hoffset = -1truecm
\voffset = -2truecm


\title{\large\bf
Three-Quark Bethe-Salpeter Vertex Function Under Pairwise 
Gluon-Exchange-Like Interaction : Application To n-p Mass Difference
}

\author{
{\normalsize\bf
Anju Sharma
}\\ 
{\normalsize National Inst. for Adv. Studies, I.I.Sc.Campus, Bangalore-560012, 
India}\and
{\normalsize\bf A.N.Mitra\thanks{e.mail: (1) insa@giasdl01.vsnl.net.in
(subj:a.n.mitra); (2) csec@doe.ernet.in(subj:a.n.mitra)}}\\
\normalsize 244 Tagore Park, Delhi-110009, India 
}

\date{13th June 1997}

\maketitle

\begin{abstract}
	A $qqq$ BSE formalism based on an input 4-fermion Lagrangian of 
`current' $u,d$ quarks interacting pairwise via a gluon-exchange-like 
propagator in its {\it non-perturbative} regime, is employed for the  
construction of a relativistic qqq-wave function  under the Covariant 
Instantaneity Ansatz (CIA). The chiral invariance of the input Lagrangian 
is automatically ensured by the vector character of the gluonic propagator,
while the `constituent' masses are the low momentum limits of the dynamical
mass function $m(p)$ generated by the standard mechanism of $DB{\chi}S$ in 
the solution of the Schwinger Dyson Equation (SDE). The CIA gives an 
exact reduction of the BSE to a 3D form which is appropriate for baryon 
spectroscopy, while the reconstructed 4D form identifies the hadron
quark vertex function as the key ingredient for evaluating transition 
amplitudes via quark-loop integrals. In this paper the second stage of this 
`two-tier' BSE formalism is extended from the 4D $q{\bar q}$-meson to the 
4D $qqq$-baryon vertex reconstruction through a reversal of steps offered 
by the CIA structure. As a first application of this 4D $qqq$ wave function, 
we evaluate the quark loop integrals for the neutron (n) - proton (p) mass 
difference which receives contributions from two sources : i) the strong 
SU(2) effect arising from the $u-d$ mass difference (4 MeV); ii) the e.m. 
effect of the respective quark charges. The resultant $n-p$ difference works 
out at 1.28 MeV (vs. 1.29 expt), with two free parameters $C_0,\omega_0$  
characterizing the infrared structure of the gluonic, which have been 
precalibrated from a common fit to $q{\bar q}$ and $qqq$ spectra as well as 
several other observable quark loop integrals. A formal derivation, based 
on Green's function techniques for 3 spinless quarks, of the CIA structure 
of the 4D $qqq$-baryon vertex function as employed in the text, is 
given for completeness in Appendix B. 
\par

	PACS : 11.10 st ; 12.35 Ht ; 12.70 + q

\end{abstract}

\newpage

\section{Introduction: Unified BS Dynamics of 2- and 3-Quark Hadrons With 
3D Kernel Support}

	Soon after the advent of the Faddeev theory [1], the 
{\it relativistic} 3-body problem [2] attracted instant attention as a 
non-trivial {\it dynamical} problem, as distinct from earlier ``kinematical" 
attempts [3] at a relativistic formulation of its wave function.
In this respect the relativistic 3-baryon problem had been more of academic 
than practical interest (until the `pion' got involved as a key ingredient), 
but the  situation changed qualitatively when this 3-body problem started 
being viewed at the quark level. Looking back after 25 years it appears that 
the first serious attempt in this direction was made by Feynman et al [4]  
who  gave  a unified formulation of both the $q{\bar q}$ (meson) and $qqq$ 
(baryon) problems under a common dynamical framework, bringing out  rather 
sharply  an  underlying duality between these two systems  which in turn 
signifies a more basic duality between a $qq$ diquark [5] and a ${\bar q}$ 
antiquark. Indeed the diquark description is quite compact and adequate 
for many practical purposes involving the baryon, but the more microscopic 
$qqq$ description which brings out the fuller permutation ($S_3$) symmetry in 
the baryon is necessary for the actual details of a full-fledged dynamical 
treatment [4].
\par
	Although the FKR theory [4] marked the first step in this direction, 
it suffered from an inadequate treatment of the time-like d.o.f. which 
showed up in several ways. The latter has by itself a long history in terms 
of attempts at formulating the Bethe-Salpeter Equation (BSE) for $q{\bar q}$ 
$qqq$ systems throughout the Seventies under the Instantaneous Approximation 
(IA), as has been reviewed elsewhere [6]. The normalization of the BS wave 
function (gaussian) has been given by Tomozawa [7] under some special
assumptions which however are not general enough to be adapted to any 
broader dynamical BSE framework going beyond the IA. 
\par
	For several years we have been involved with a certain formulation 
of BS-dynamics for both $q{\bar q}$ and $qqq$ systems within a common 
{\it unified} framework (to emphasize their underlying duality), designed  
to address their spectroscopy on the one hand, {\it and} a self-consistent 
treatment of various quark-loop amplitudes in terms of their respective 
hadron-quark vertex functions on the other. The ``spectroscopy" aspects are 
addressed through the 3D reductions of the $q{\bar q}$ BSE [8] and the $qqq$
BSE [9], to compare with observed O(3) spectra [10], while the loop-integral
aspects of transition amplitudes show up through the reconstructed  vertex 
functions of the 4D BSE [11]. An {\it {exact interconnection}} between the two 
forms was achieved through the ansatz of a 3D support defined {\it covariantly}
in the BS kernels for the $q{\bar q}$ [12] and $qqq$ [13] systems. 
\par
	The ``Covariant Instaneity Ansatz" (CIA) [12] which has been at the 
root of this ``two-tier" philosophy, and is also supported by other 
considerations [14] based on the Markov-Yukawa transversality condition [15], 
seems to offer a possible Lorentz covariant way to reconcile the apparently 
conflicting demands of 3D spectroscopy [10] with the 4D structure of 
quark-loop amplitudes. The effectiveness of the CIA in giving a concrete 
shape to such a ``two-tier" philosophy of spectra-cum-loop integrals was 
summarised in a  semi-review [13] in the form of appropriate BSE's for 
$q{\bar q}$ and $qqq$ systems with vector-type kernels [6] with 3D support, 
albeit with slight modifications [6] in the respective BSE structures to 
facilitate greater `manoeuvreability', in the spirit of similar efforts [16] 
in the past. Further, the observed spectroscopy [10] is well satisfied on 
both $q{\bar q}$ [8] and $qqq$ [9] sectors with a {\it {common set}} of 
parameters for the respective kernels (the $qq$ kernel has just half the 
strength of the $q{\bar q}$ kernel due to color effects), so that the 
respective vertex functions are entirely determined within this formalism.
\par
	The other aspect of this `two-tier' formalism concerns the crucial
property of chiral symmetry {\it and} its dynamical breaking. The first part
(chiral symmetry) is ensured without extra charge by the {\it vector} 
character of the kernel that had been present all along in this program 
[6,11], since the BS-kernel is a direct reflection of an effective 4-fermion 
term in the input Lagrangian. Indeed the vector type character of the latter 
lends  a natural gluon exchange flavour to such a pairwise interaction among 
`current' (almost massless) $u,d$ quarks at the Lagrangian level. This 
structure is quite general [17], and can be adapted to the QCD requirements 
on the gluonic propagator involved in the pairwise interaction kernel. Of 
this, the perturbative part (which is well understood) is quite explicit, 
but the {\it non-perturbative} (infrared) part is not yet derivable from 
formal QCD premises. It can nevertheless be simulated in a sufficiently 
realistic manner at the phenomenological level [6,19], so as to satisfy the 
standard constraints of confinement as well as explicit QCD features [20] in 
terms of a basically 3D BSE kernel structure. 
\par
	The second part, viz., dynamical breaking of chiral symmetry 
($DB{\chi}S$) is implemented via the Nambu-Jonalasino mechanism [21] whose 
full-fledged form amounts to adopting the `non-trivial' solution of the 
Schwinger-Dyson Equation (SDE) derived from a given input, chirally symmetric 
Lagrangian with current quarks. A mass function $m(p)$ [17,18] is thus 
generated whose low-momentum value may be identified with the bulk of the 
`constituent' mass ($m_q$) of the $u,d$ quarks. This accords with Politzer 
additivity [22], viz., $m_q = m(0) + m_c$; where $m_c$, the current mass, 
is small. This was also shown in the context of a BSE-cum-SDE treatment [23] 
within the CIA formalism [12]. Thus formally the BS-kernel may be regarded as 
a {\it non-perturbative} gluon propagator [23] in a BSE framework involving 
the dynamical/constituent mass [19, 21-24] in the quark propagator.  
\par
	To recapitulate, the CIA which gives an exact interconnection between
the 3D and 4D forms of the BSE, provides a unified view of 2- and 3-quark 
hadrons, its 3D reduction being meant for spectroscopy [8-9], and the 
reconstructed 4D form [12,13] for identifying the respective hadron-quark 
vertex functions as the key ingredients for evaluating 4D quark-loop integrals.
The formalism stems from a strongly QCD-motivated Lagrangian with 
{\it current} quarks whose pairwise interaction is mediated by a gluonic 
propagator in its non-perturbative regime. The QCD feature of chiral symmetry 
is ensured by the vector nature of this interaction, while its dynamical 
breaking is the result of a non-trivial solution of the SDE [17,23]. Thus, 
unlike in conventional potential models [25], the constituent mass so generated
is {\it not} a phenomenological artefact, but the result of a self-consistent 
solution of the SDE [17, 19, 23], wherein the (constituent) mass normally
employed for spectroscopy [8-9] {\it matches} with the output dynamical mass 
at low momentum [23]. Thus there are only two genuine input parameters $C_0$, 
$\omega_0$, that characterize the (phenomenological) structure of the 
non-perturbative gluon propagator which serves for both the 2- and 3-quark 
spectra in a unified fashion [8-9]. In this formalism, these two constants 
play a role somewhat similar to that of the (input) `condensates' in the 
theory of QCD sum rules [26].
\par      	    		 
	Before proceeding further, let us pause to compare this approach with
other dynamical methods, e.g., chiral perturbation theory [27] which has 
more explicit QCD features, albeit in the perturbative regime, leading to
expansions in the momenta. This is a powerful theoretical approach employing 
the (chiral) symmetry of QCD; its essential parameters are the current quark 
masses, and the method works very efficiently where its premises are 
logically applicable. Thus it predicts the {\it {ground state}} spectra of
light quark hadrons, including their mass splittings due to strong and e.m.
breaking of SU(2), but {\it not} the spectra of L-excited hadrons. The latter
on the other hand demand a ``closed form" approach to incorporate the ``soft"
{\it {off-shell}} effects which in turn require a non-trivial handle on the
{\it infrared} (non-perturbative) part of the gluonic propagator, something 
which the present state of the QCD art does not yet provide. Thus one needs a 
phenomenological input even in standard BSE-SDE approaches [18], as discussed 
elsewhere [23]. The chiral perturbation theory [27] also lacks this vital 
ingredient, as seen from the absence of form factors in its `point' 
Lagrangians [27] with at most derivative terms. This shows up, e.g., through 
its inability to predict L-excited spectra, and finer aspects (such as 
convergence) of 4D quark-loop integrals which depend crucially on these 
``off-shell" features. {\it Physically} this amounts to the absence of a 
`confinement scale' which governs these form factors. In other BSE-cum-SDE 
approaches [17-19], including the present `two-tier' CIA formalism [12,23], 
this `scale' is an integral part of the structure of the non-perturbative part 
of the gluon propagator [19,23], with a built-in QCD feature of chiral 
symmetry and its dynamical breaking through the non-trivial solution of the 
SDE [17,19,23]. This not only facilitates the prediction of L-excited spectra 
[19, 8-9] but also provides a form factor for the hadron-quark vertex function
which greatly enhances its applicability to various 4D quark-loop integrals; 
see [12, 23-24, 28].    
\par
	After this clarification on the philosophy of this two-tier BSE 
approach, vis-a-vis some others [26,27], we may now state the objective of 
the present paper: A typical application of the 4D baryon-$qqq$ wave function
reconstructed [13] from the 3D $qqq$ BSE, as a 3-body generalization of the  
corresponding $q{\bar q}$-meson problem [12]. Unlike the 2-body case, however, 
where the steps are exactly reversible [12], such reconstruction in the 3-body
case involves a loss of information on the 4D Hilbert space for a 3-body 
system, so that the reversal of steps is in principle not unique, and requires 
a 1D $\delta$-function to fill up the information gap between 3D and 4D 
Hilbert space which may be directly attributed to the CIA ansatz of a 3D 
support to the pairwise kernel. The 2-body case just escapes this pathology 
as it represents a sort of degenerate situation, but the price of a 3D kernel 
support must show up in a reconstruction of the 4D BSE from its reduced 3D 
form in any $(n > 2)$-body problem [29]. A plausible `CIA' structure for the 
4D $qqq$ wave function was suggested in [13] in a semi-intuitive fashion, 
but a more formal mathematical basis has since been found [29] through the 
use of Green's function techniques, so that the reconstructed 4D form reduces 
exactly to the (known) 3D form as a consistency check [29]. The final result, 
which is almost the same as the earlier conjecture, eq.(5.15) of [13], except 
for a constant that does not affect the normalization, contains a 1D 
$\delta$-function corresponding to the on-shell propagation of the spectator 
between two successive vertex points. As explained in detail in [29], this 
1D $\delta$-function must not be confused with any signature of 
``non-connectedness" in the 3-body wave function [30], since the 3D form is 
{\it {fully connected}}. A better analogy is to the `scattering length 
approximation' to the $n-p$ interaction, characterized by the appearance
of a (Fermi-type) $\delta$-function potential, in estimating the effect of 
chemical binding on the scattering of neutrons by a hydrogen molecule [31]. 
In any case the 1D $\delta$-function appearing in this structure is entirely 
innocuous since it gets integrated out in any {\it physical} (quark loop) 
amplitude including the BS normalization (see Sec.2 below).
\par
 
	The application chosen for the baryon-$qqq$ vertex function is to  
the n-p mass difference, on closely parallel lines to the meson case [28].
To racall the {\it physics} of the n-p mass difference, this quantity receives
contributions of {\it opposite} signs from two distinct sources : 
i) a positive one from the strong SU(2) $d-u$ mass difference; \\
ii) a negative one from e.m. splittings.
\par 
	For the sake of completeness, Appendix B of this paper gives 
the main steps of the derivation [29] for the 4D structure, eq.(5.15) of
[13], of the baryon-$qqq$ vertex function by the Green's function method 
for 3 spinless quarks, to be employed in this paper. In Sec.2 we collect
the various pieces of this quantity with the inclusion of the spin and 
isospin d.o.f., on the lines of [32]. Thus equipped, we outline the main 
steps leading to an explicit evaluation of the normalization integral, using 
Feynman diagrams shown in figs.1(a.b,c). A {\it complex} basis [9, 33, 34] 
for 3D momentum variables facilitates the evaluation of the resulting 
$3D\times 3D$ integrals, {\it after} the time-like momenta have been 
eliminated by `pole' integrations on identical lines to the corresponding 
$q{\bar q}$ problem [12,23,24]. In Sec.3 we evaluate the `shift' in the 
nucleon mass due to strong SU(2) breaking, by inserting a mass shift operator  
$- \delta m {\tau_3}^{(i)}/2$  in place of $i{\hat \gamma}_\mu e_i$ at each of 
the corresponding $\gamma$- vertices of figs.1(a,b,c), as shown in figs. 
2.(a,b,c). Here $\delta m$ = 4 MeV is the `standard' d-u mass difference 
[24,28] taken as the basic input. The condensate contribution is neglected as 
it was found to be negligible from a similar calculation of the SU(2) mass 
splittings in pseudoscalar mesons [28]. Sec.4 sketches the evaluation of the 
e.m. contribution in accordance with the diagrams of fig.3(a,b,c),  while the 
details of the approximations employed are collected in Appendix A. Sec.5 
summarises our findings and conclusions vis-a-vis other methods.
\section{Normalization of the Baryon-$qqq$ Vertex Function}
\par
	To outline the structure of the baryon-$qqq$ vertex function 
from a CIA-governed BSE [12-13], we shall generally follow the notation, 
normalization and phase convention for the various symbols as given in [13], 
but adapted to the {\it equal} mass kinematics $(m_1 = m_2 = m_3 = m_q)$. 
The SU(2) mass difference $\delta m$ $(\approx 4 MeV)$ between $d$ and $u$
quarks will be taken into account only through a 2-point vertex 
$[- \delta m {\tau_3}^{(i)} /2]$ inserted in the quark propagators in figs.2 
(in place of $i\gamma_\mu e_i$  for a photon), but not in the structure of 
the vertex function. The vertex function is written in {\it three} pieces in 
each of which {\it one} quark plays the role of the `spectator' by turn. For 
the spin structure (not given in [13]) we employ the convention of [3] which 
was extended in [32] to incorporate the $S_3$-symmetry for the spin-cum-
isospin structure in the Verde [35] notation [36]. The full 4D BS wave 
function $\Psi$ reads as [13,32,33] : 
\setcounter{equation}{0}   
\renewcommand{\theequation}{2.\arabic{equation}}
\begin{equation} 
\Psi \Delta_1 \Delta_2 \Delta_3 = (\Gamma_1 + \Gamma_2 + \Gamma_3) \times
  [\chi' \phi' + \chi" \phi"]/{\sqrt 2};
\end{equation}
\begin{equation}
\Delta_i = {m_q}^2 + {p_i}^2 ; \quad (i = 1,2,3).
\end{equation}
Here $\chi'$ and $\chi"$ are the relativistic ``spin" wave functions 
in a 2-component mixed symmetric $S_3$  basis which for a {\bf 56} baryon 
go with the associated isospin functions $\phi'$ and $\phi"$ respectively. 
These are given by [3,17] : 
\begin{equation}
[\chi']_{\beta\gamma;\alpha} = [(M-i\gamma.P) i\gamma_5 C/{\sqrt 2}]_
{\beta\gamma} \times U(P)_\alpha/(2M)
\end{equation}
\begin{equation}
[\chi"]_{\beta\gamma;\alpha} = [(M-i\gamma.P) \gamma_\mu C/{\sqrt 6}]_
{\beta\gamma} \times {i\gamma_5 \gamma_\mu U(P)}_\alpha/(2M)
\end{equation}
in a spinorial basis [3,32] in which the index $\alpha$ refers to the 
`active' quark (interacting with an external photon line, fig.1), 
while $\beta,\gamma$ characterize the other two, with the further convention 
that $\gamma$ refers to the ``spectator" in a given diagram, fig.(1).
The `hat' on  $\gamma$ signifies its perpendicularity to $P_\mu$, viz., 
${\hat \gamma}.P = 0$. The notations in eqs.(2.3-4) are standard, with a 
common Dirac basis for the entire structure, and `C' is the charge 
conjugation operator for quark $\# 3$ in a 23-grouping [3,32]. $P_\mu$
is the baryon 4-momentum, $U(P)$  is  its  spinor  representation, 
and $(M-i\gamma.P)/(2M)$ its energy projection operator [3,32].  Further,
because of the full $S_3$-symmetry of the last factor in (2.1), the
$(1,2,3)$ indices can be permuted as needed for the diagram on hand. 
Thus in fig.1a, $\# 1(\alpha)$ interacts with the photon ; $\# 2(\beta)$ is 
the quark which has had a `last' $qq$-interaction with $\# 1(\alpha)$  before 
emerging from the hadronic `blob', while $\# 3(\gamma)$ is the spectator 
[32]. In fig.1b, the roles of $\# 1$ and $\# 2$ are reversed so that, of the 
two `active quarks' $\# 1$ and $\# 2$, $\# 2(\alpha)$ now interacts with the 
photon, $\# 1(\beta)$ has had the last $qq$- interaction with $\# 2(\alpha)$,
while $\# 3(\gamma)$ still remains the `spectator'. These roles are cyclically 
permuted, with two more such pairs of diagrams, fig.1c), to give an 
identical chance to each of the quarks in turn [32]. Thus there are 3 such 
pairs of diagrams, of which only one pair is shown. An identical consideration
applies to figs.2(a,b) with $i\gamma_\mu e_i$  replaced by 
$(-\delta m \tau_3 /2)$ consistently. The spatial vertex functions $\Gamma_i$
are given for i = 3 by [13] :
\begin{equation}
\Gamma_3 = N_B [D_{12} \phi /{2i\pi}]
\times {\sqrt {[2\pi\delta(\Delta_3).\Delta_3]}}
\end{equation} 
where $\phi$ is the full, {\it connected} $qqq$ wave function in 3D form, and 
$D_{12}$ is the 3D denominator function of the (12) subsystem . The second 
factor represents the effect of the spectator [13] whose {\it inverse}                
propagator ${D_F}^{-1}(p_3)$ {\it {off the mass shell}} is just $\Delta_3$, 
eq.(2.2). The main steps leading to this unorthodox structure which has been 
derived recently via the techniques of Green's functions [29], are sketched 
for completeness in Appendix B. As already noted in Sec.1, and again explained
in Appendix B, its peculiar singularity structure in the form of a  
``square-root" of a 1D $\delta$-function stems from the CIA ansatz of a 3D 
support to the pairwise interaction kernel, but it is quite harmless as the 
former will appear in a {\it linear} form in the transition amplitude 
corresponding to any Feynman diagram as in figs.1-2. The complete expressions 
for $D_{12}$ and $\phi$ are given for the equal mass case (with $\# 3$ as 
spectator) by [13] (see also [9]): 
\begin{equation}
D_{12} = \Delta_{12} (M-\omega_3); \quad 
\Delta_{12} = 2 {\omega_{12}}^2 - M^2 (1-\nu_3)^2/2
\end{equation}
\begin{equation}
{\omega_{12}}^2 = {m_q}^2 + {{\hat q}_{12}}^2; \quad
2{{\hat q}_{12}}^\mu = {{\hat p}_1}^\mu - {{\hat p}_2}^\mu
\end{equation}
\begin{equation}
{\phi} = \ e^{-({{\hat p}_1}^2 + {{\hat p}_2}^2 + {{\hat p}_3}^2)/{2\beta^2}}
     \equiv \ e^{-\rho/{3\beta^2}}
\end{equation}
(see further below for the definition of $\rho$).
\begin{equation}
{{\hat p}_i}^\mu = {p_i}^\mu + p_i.P P_\mu/M^2; \quad
{{\hat p}_1}^\mu + {{\hat p}_2}^\mu + {{\hat p}_3}^\mu = 0
\end{equation}
\begin{equation}
{\omega_i}^2 = {m_q}^2 + {{\hat p}_i}^2; \quad \nu_3 = \omega_3/M (on shell)
\end{equation}
The $\beta$-parameter is defined sequentially by [8,9]:
\begin{equation}
\beta^4 = {4 \over 9}M {\omega_0}^2 {\bar \alpha}_s(1-m_q/M)^2 (M-<\omega>);
\quad {<\omega>}^2 = {m_q}^2 + 3\beta^2/8
\end{equation}
\begin{equation}
{{\bar \alpha}_s}^{-1} = {\alpha_s}^{-1} - {2M C_0}{(1-m_q/M)^2 
\over{M-<\omega>}}; 
\end{equation}
\begin{equation}
{6\pi \over{\alpha_s}} = 29 {\ln {(M-<\omega>) \over {\Lambda_{QCD}}}};
\end{equation}
\begin{equation}
\Lambda_{QCD} = 200 MeV;\quad \omega_0 = 158 MeV;\quad C_0 = 0.29
\end{equation}   
The normalization $N_B$, eq.(2.5), is given in accordance with the Feynman 
diagrams 1(a,b) by the 4D integral ( c.f.[32]) :
\begin{eqnarray}
iP_\mu/M & = & \sum_{123} \int d^4q_{12} d^4p_3{{{\Gamma_3}^*\Gamma_3}
\over{2\Delta_2 \Delta_3}}[<\phi'|(23)'(1)'_\mu |\phi'> + 
{1\over 3}<\phi"|(23)"_{\nu\lambda}(1)"_{\nu\lambda;\mu}|\phi">] \nonumber \\
         &   & + (1 \Leftrightarrow 2)
\end{eqnarray}
where the matrix element for fig.1a is organized as a product of two 
spin-factors : a `23-element' expressed as a Dirac trace over the indices 
$\beta,\gamma$; and a `1-element' (with suppressed index $\alpha$).
The associated isospin functions $\phi$ are shown according to (2.1).
The contribution of fig.1b is shown symbolically by $1 \Leftrightarrow 2$, 
while $\sum_{123}$ indicates the sum over all the 3 pairs cyclically. In       
representing eq.(2.12) we have dropped `cross-terms' like 
${\Gamma_i}^*\Gamma_j$, where $i \neq j$, since the presence of a 
$\sqrt \delta$-function in each $\Gamma_i$  ensures that a simultaneous 
`on-shell' energy conservation of $i \neq j$ spectators is not possible [32]. 
The various pieces of the matrix elements in (2.14) which can be read off 
from fig.1a in terms of the spin functions (2.3-4) are as follows:
\begin{equation}
(1)"_{\nu\lambda;\mu} = {\bar U}(P)S_F(p_1)i\gamma_\mu e_1 S_F(p_1) U(P)
\end{equation}
\begin{equation}
i{S_F}^{-1}(p) = m_q + i\gamma.p
\end{equation}
\begin{equation}
(1)"_{\nu\lambda;\mu} = {\bar U}(P) i{\hat \gamma}_\nu \gamma_5 S_F(p_1)
i\gamma_\mu e_1 S_F(p_1) i\gamma_5 {\hat \gamma}_\lambda U(P);
\end{equation}
\begin{equation}
(23)' = Tr[C^{-1}\gamma_5(M-i\gamma.P)(m_q-i\gamma.p_2)(M-i\gamma.P)
\gamma_5(m_q+i\gamma.p_3)C]/{8M^2}
\end{equation}
\begin{equation}
(23)"_{\nu\lambda} = Tr[C^{-1}{\hat \gamma}_\nu(M-i\gamma.P)(m_q-i\gamma.p_2)
(M-i\gamma.P){\hat \gamma}_\lambda(m_q+i\gamma.p_3)C]/{8M^2}
\end{equation}								    
The `strength' $e_i$  of the (zero-momentum) `photon' coupling to              
the quark line $p_i$  can be chosen in several ways [37]. We take                   
here the simplest possibility, viz., $e_i = 1/3$ each. The isospin                                  
matrix element is first eliminated according to [38]:
\begin{equation}
<\phi'|1|\phi'> = <\phi"|1|\phi"> = 1
\end{equation}
\begin{equation}
<\phi'|{\tau_3}^{(1)}|\phi'> = -3<\phi"{\tau_3}^{(1)}|\phi"> = <\tau_3>_{(p,n)}
\end{equation}
Eq.(2.20) suffices for (2.14), while (2.21) will be needed for                                     
the u-d mass difference operator $-\delta m{\tau_3}^{(1)}/2$ ; see Sec.3. 
Next, the evaluation of the traces in (2.15-18) is straightforward,
after noting that\ (2.15-16), {\it after} spin-averaging, are expressible 
as traces. The results are 
\begin{equation}
(23)' \theta_{\nu\lambda} = (23)"_{\nu\lambda} 
 = (m_q + M\nu_2)(m_q + M\nu_3) \theta_{\nu\lambda}
\end{equation}
\begin{equation}
(1)'_\mu \theta_{\nu\lambda} = (1)"_{\nu\lambda;\mu} 
 = [2M\nu_1(m_q + M \nu_1) + \Delta_1]\theta_{\nu\lambda}P_\mu/(M{\Delta_1}^2)
\end{equation}
where $\theta$ is a covariant Kronecker delta w.r.t. $P_\mu$, viz., 
\begin{equation}
\theta_{\nu\lambda} \equiv \theta_{\nu\lambda} 
 = \delta_{\nu\lambda} - P_\nu P_\lambda/P^2 ; \quad (P^2 = -M^2)
\end{equation}                                               
Collecting all these results and simplifying we get 
\begin{equation}
{N_B}^{-2} = \sum_{123} \int d^3{\hat p}_3 {(m_q + \omega_3)\over{2\omega_3}}
\times \int d^3{\hat q}_{12} {D_{12}}^2 {\phi}^2 [e_1 I_1 + e_2 I_2]
\end{equation}
\begin{equation}
2i\pi I_1 = \int Md\sigma_{12}[2M\nu_1(m_q + M\nu_1) + \Delta_1]/
(M{\Delta_1}^2 \Delta_2)
\end{equation}
where we have ``cashed" the $\delta(\Delta_3)$-function arising from 
$|\Gamma_3|^2$ against the time-like component of $d^4p_3$, and used 
the results 
\begin{equation}
d^4q_{12} = d^3{\hat q}_{12} M d\sigma_{12}; \quad 
\nu_{1,2} = (1 - \nu_3) \pm \sigma_{12}
\end{equation}
The integration over $d\sigma_{12}$ involves single and double poles arising 
from the propagators ${\Delta_{1,2}}^{-1}$ in (2.26), while the value of 
$\nu_3$ is taken `on-shell' at $\omega_3 /M$ after the $\delta(\Delta_3)$-
function has been cashed. The result of a basic $\sigma_{12}$-integration is  
\begin{equation}
\int M d\sigma_{12} {\Delta_1}^{-1} {\Delta_2}^{-1} = 2i\pi/D_{12}
\end{equation}
from which others can be deduced by differentiation under {\it unequal} 
mass kinematics, or directly through a `double pole' integration. The net 
result for $I_1 + I_2$, eq.(2.26), is given in eq.(2.41) below. Further, 
the individual terms of the summation $\sum_{123}$ in (2.25) are fixed by 
the values chosen for $e_i$ (which need not be specified in advance, as they 
can be adapted to other conventions too [37]; see Sec.3).
\par
 
	The integration in (2.25) can be considerably simplified in  a 
complex basis [9,33] defined (in momentum space) by :
\begin{equation}
{\sqrt 2} z_i = \xi_i + i\eta_i; \quad {\sqrt 2} {z_i}^* = \xi_i - \eta_i;
\end{equation}
\begin{equation}
{\sqrt 3}\xi_i = p_{1i} - p_{2i}; \quad 3\eta_i = -2p_{3i} + p_{1i} + p_{2i};
\end{equation}                                         
where we now employ the alternative notation $p_{1i}$ for ${\hat p}_1^\mu$, 
in view of its basically 3D content. In terms of $z_i$  and ${z_i}^*$, the                     
6D integration in (2.25) is expressed as 
\begin{equation}
d^3{\hat p}_3 d^3{\hat q}_{12} = ({\sqrt 3}/2)^3 d^3 \xi d^3 \eta = d^3z d^3z^*
\end{equation} 
The further representation [9,33] 
\begin{equation}
d^3z d^3z^* = ({dz_+}{dz_-}^*) ({dz_-}{dz_+}^*)({dz_3}{dz_3}^*)
\end{equation}
where
\begin{equation}
{\sqrt 2}z_+ = R_1 \ e^{i\theta_1};
\quad {\sqrt 2}{z_-}^* = R_1 \ e^{-i\theta_1}
\end{equation}
\begin{equation}
{\sqrt 2}z_- = R_2 \ e^{i\theta_2};
\quad {\sqrt 2}{z_+}^* = R_2 \ e^{-i\theta_2}
\end{equation}
\begin{equation}
{\sqrt 2}z_3 = R_3 \ e^{i\theta_3};
\quad {\sqrt 2}{z_3}^* = R_3 \ e^{-i\theta_3}
\end{equation}
reduces the 6D integration (2.32)  merely to 
$\pi^3 d{R_1}^2 d{R_2}^2 d{R_3}^2$, since the $\theta_i$ -variables 
({\it not} Euler angles!) are not involved in the integrands encountered, and 
just sum up to $(2\pi)^3$.  The positive variables $R_i$, (i = 1,2,3), are 
related to the $\xi_i,\eta_i$ variables by 
\begin{equation}
\rho \equiv R_1^2 + R_2^2 + R_3^2 = \xi^2 + \eta^2 = 2 z_i {z_i}^*
\end{equation}
\par
                                         
	To convert the variables $\omega_i$ that appear in the integrals 
(2.28) in terms of the $R_{1,2,3}$  variables is a straightforward but 
tedious process which can be somewhat simplified in terms of the        
intermediate variables $\xi^2 - \eta^2 $ and $2 \xi.\eta$ which form a [2,1] 
representation [35] of $S_3$-symmetry at the `quadratic' level. Now because 
of the full $S_3$ -symmetry of the 6D integral (2.32), together with the 
(fortunate) circumstance of equal mass quarks in the problem on hand, the 
integrand as a whole is $S_3$ -symmetric which permits the following 
simplification: Each of the quantities ${\hat p}_i^2$  and ${\hat q}_i^2$   
inside (2.32) can be expanded as 
\begin{equation}
{{\hat p}_{1,2}}^2 = \rho/2 + (\xi^2 - \eta^2)/4 \pm {\sqrt 3}\xi/\eta/2;
\quad {{\hat p}_3}^2 = \rho/2 - (\xi^2 - \eta^2)/2
\end{equation}
\begin{equation}
{{\hat q}_{12}}^2 = 3\xi^2/4 = \rho/2 + (\xi^2 - \eta^2)/4
\end{equation} 
In all these terms the principal quantity is $\rho/2$, while the 
resultant effects of the mixed-symmetric corrections will show up 
only in the {\it fourth} order, etc. In the present case of equal mass 
kinematics it is a good approximation to neglect the latter terms, as has 
also been found for the qqq mass spectral results [9], so that all 
quantities are expressed in terms of $\rho$ only :
\begin{equation}
\omega_{1,2,3} \approx \omega_{12} \approx \omega_\rho;
\quad {\omega_\rho}^2 \equiv m_q^2 + \rho/2
\end{equation}
\begin{equation}
D_{12} \approx 2(M-\omega_\rho)[\omega_\rho^2 - (M-\omega_\rho)^2/4].
\end{equation}
The rest of the integration is expressed entirely in terms of the 
$\rho$-variable, with the resultant 6D measure given by 
\begin{equation}
\int d^3{\hat p}_3 d^3{\hat q}_{12} F(\rho)
= (\pi{\sqrt 3}/2)^3 \int \rho^2 d\rho/2 F(\rho)
\end{equation}  
These considerations suffice for evaluating the integrals $I_1$                                                               
and $I_2$   whose resultant value is now given for $e_i = 1/3$ by :
\begin{equation}
D_{12}^2 (I_1 + I_2) = [m_q^2 + m_q(M-\omega_\rho) + (M-\omega_\rho)^2/4]
\times (M-\omega_\rho)^3/\omega_\rho + D_{12} (2m_q + M - \omega_\rho)
\end{equation}
Substitution in (2.25) yields $N_B$ directly. The numerical values 
are given collectvely at the end of Sec.4.
\section{Strong SU(2) Mass Difference for the Nucleon}
\par

	This calculation is on almost identical lines to Sec.2, except                                  
for the substitution $ie_1 \gamma_\mu$ to $-\delta m {\tau_3}^{(1)}/2$ in 
figs.1(a,b) to give figs.2(a,b) which represent the effect of insertion of 
a 2-point vertex in a quark line. Indeed we can directly start from the 
counterpart of eq.(2.15) which gives the `strong' mass shift as:
\setcounter{equation}{0}
\renewcommand{\theequation}{3.\arabic{equation}}
\begin{eqnarray}
i\delta M_{st} & = & \sum_{123} \int d^4q_{12}d^4p_3 {{\Gamma_3}^* \Gamma_3
\over {2\Delta_2\Delta_3}}\times [<\phi'|(23)'(1)'|\phi'> + {1\over 3}<\phi"|{(23)"}_{\nu\lambda}{(1)"}_
{\nu\lambda}|\phi">]  \nonumber \\
               &   & + (1 \Leftrightarrow 2)
\end{eqnarray}
where we have now employed eq.(2.21) for the isospin factors, and the 
counterparts of (2.16) and (2.17) are respectively 
\begin{equation}
(1)' = {\bar U}(P)S_F(p_1)[-\delta m {{\tau_3}^{(1)}}/2]S_F(p_1) U(P);
\end{equation}
\begin{equation}
(1)"_{\nu\lambda} = {\bar U}(P)i{\hat \gamma}_\nu \gamma_5 S_F(p_1)
[-\delta m {{\tau_3}^{(1)}}/2]S_F(p_1) i\gamma_5 {\hat \gamma}_\lambda U(P)
\end{equation}  
while the definitions (2.18) and (2.19) remain unaltered. As a result, 
eq.(2.22) remains valid, while the counterpart of (2.23) becomes 
\begin{equation}
(1)'\theta_{\nu\lambda} = -3 (1)"_{\nu\lambda} 
= [2m_q(m_q + M \nu_1) - \Delta_1]\theta_{\nu\lambda}(-\delta m/2)/\Delta_1^2
\end{equation}
Carrying out the $d\sigma_{12}$-integration, the result for $\delta M_{st}$ 
is now given by the counterpart of (2.25), viz.,  
\begin{equation}
\delta M_{st} = 3N_B^2 \int d^3{\hat p}_3{(m_q + \omega_3)\over {2\omega_3}}
\times \int d^3{\hat q}_{12} {D_{12}}^2 \phi^2 [J_1 + J_2](-\delta m\tau_3/6)
\end{equation}
in the form of an isospin operator ``$\tau_3$" for the {\it nucleon}, where                                        
we have represented the effect of $\sum_{123}$ by a factor of ``3", and 
\begin{eqnarray}
D_{12}^2 [J_1 + J_2] & = & {m_q\over {\omega_\rho}} [(m_q + {1\over 2}
(M-\omega_\rho))(M-\omega_\rho)[2\omega_\rho^2 + m_q(M-\omega_\rho)] \nonumber \\
                     &   & +(M-\omega_\rho)^2[m_q + (M-\omega_\rho)/2]^2 
+ \Delta_{12}(m_q^2-\omega_\rho^2) \nonumber \\  
                     &   & + (m-\omega_\rho)(m_q + (M-\omega_\rho)/2)(\omega_\rho^2 + m_q(M-\omega_\rho)/2) \nonumber \\        
                     &   & + \Delta_{12}^2 (1 -m_q(M-\omega_\rho)/\omega_\rho^2)/2]
\end{eqnarray}
as the exact counterpart of (2.42) under the same approximation. It is seen 
from (3.5) that the {\it difference} $(n-p)$ is {\it positive}.
\section{E.M. Mass Difference for the Nucleon}
\par

	The diagrams for the e.m. mass difference are given by figs.3 
(I,II,III) for a proton ($uud$) configuration to illustrate the 
underlying topology in accordance with the roles of the `active' and 
`spectator' quarks in turn, as explained in Sec.2. In each of these diagrams, 
two internal quark lines are joined by a photon line. The e.m. vertex at 
quark $\# i$ has the strength $e[1 + 3\tau_3^{(i)}]/6$ from  which the  isospin 
matrix elements of a product of {\it two} such factors (shown for fig.3.III) 
have the forms
\setcounter{equation}{0}
\renewcommand{\theequation}{4.\arabic{equation}} 
\begin{equation}
<\phi';\phi"|{(1+3\tau_3^{(1)})/6} \times {(1+3\tau_3^{(1)})/6}|\phi';\phi">
\end{equation}
for the proton $(uud)$ cnfiguration shown in III with $\# 3$  as spectator, but 
in a basis (1;23) (which is consistent with the spin basis, eqs.(2.3-4)), 
corresponding to fig.1a, viz.[36,38]:
\begin{equation}
|\phi'> = u_1(u_2d_3-u_3d_2)/{\sqrt 2}; \quad
|\phi"> = (-2d_1u_2u_3 + u_1d_2u_3 + u_1u_2d_3)/{\sqrt 6}
\end{equation}
We note in parentheses that in fig.3.III, the interchange of the two `active' 
quarks $\# 1$ and $\# 2$ does {\it not} give a new configuration, {\it unlike} in 
figs.1 and 2; ((a) versus (b) configurations).
\par

	It is now easy to check that the matrix elements $< >'$  and $< >"$
of (4.1) are $1/9$ and $-1/9$ for the {\it proton} configuration. After doing 
the corresponding neutron case, the two results may be combined in the single 
operator forms [38]:
\begin{equation}
<.>' = (1 + 3\tau_3)/36; \quad <.>" = (1 - 5\tau_3)/36
\end{equation}
where $\tau_3$ is the isospin operator for the nucleon as a whole [see 
eq.(3.5)], to be sandwiched between the neutron and proton states. The  
resultant isospin factor is then 
\begin{equation}
e^2[<.>' + <.>"]/2 = e^2(1-\tau_3)/36 \Rightarrow -e^2\tau_3/36
\end{equation}                                      
After this book-keeping on the charge factors we can drop the isospin d.o.f. 
$|\phi>$ from the $qqq$ wave function and, on the basis of the equality of 
the $(.)'$ and $(.)"$ contributions (2.22-23) for the spin matrix elements, 
it is enough to work with the $(.)'$ type to represent the full effect. 
Collecting these details, the net isospin contribution to the e.m. $(n-p)$  
mass difference is just $e^2/18$, which (of course) comes out with the 
{\it correct} (negative) sign in the resultant e.m.contribution to the total
$n-p$ difference after all the phase factors in the orbital-cum-spin space 
have been taken into account. The complete e.m. self energy of the nucleon  
(with operator $\tau_3$), with  fig.3.III as the prototype, is now given by 
\begin{equation}
\delta M^\gamma = \sum_{123} [-e^2 \tau_3/36]/(2\pi)^4 \int d^4p_3d^4q_{12}
d^4{q_{12}}'\Gamma_3^* \Gamma_3' k^{-2} \times [23]_\mu'[1]_\mu'/
(\Delta_3 \Delta_2 \Delta_2')
\end{equation} 
where the various momentum symbols are as shown in fig.3, with the primed 
quantities referring to the vertex on the  right, but otherwise written in 
the same convention as in eqs.(2.5-10). The symbols within square brackets 
are analogous to (2.16-19):
\begin{equation}
[1]_\mu' = {\bar U}(P')S_F(p_1')i\gamma_\mu S_F(p_1)U(P); \quad (P'=P)
\end{equation}
\begin{equation}
[23]_\mu' = {Tr\over {8M^2}}[C^{-1}\gamma_5(M-i\gamma.P')(m_q-i\gamma.p_2')i\gamma_\mu
(m_q-i\gamma.p_2)(M-i\gamma.P)\gamma_5 (m_q+i\gamma.p_3) C]
\end{equation} 
And the product of (4.6) and (4.7) works out as 
\begin{eqnarray}
ME & \equiv & {(m_q + \omega_3)\over {\Delta_1 \Delta_1'}}[-(\Delta_1 + 
\Delta_1' - k^2)(\Delta_2 + \Delta_2' -k^2)/4 \nonumber \\
   &        & -(\Delta_1 + \Delta_1'-k^2)(m_q\omega_2 + m_q\omega_2'+
2\omega_2\omega_2')/2 \nonumber \\
   &        & - (\Delta_2 + \Delta_2'-k^2)(m_q\omega_1 + m_q\omega_1'+
2\omega_1\omega_1')/2 \nonumber \\
   &        & -(\Delta_1 + \Delta_2)(m_q + \omega_1')(m_q + \omega_2')/2
- (\Delta_1' + \Delta_2')(m_q + \omega_1)(m_q + \omega_2)/2 \nonumber \\
   &        & -(\Delta_1' + \Delta_2)(m_q + \omega_1)(m_q + \omega_2')/2
-(\Delta_1 + \Delta_2')(m_q + \omega_1')(m_q + \omega_2)/2 \nonumber \\
   &        & +(m_q^2 +{1\over 2}(P-p_3-k)^2)[(m_q + \omega_1)(m_q + \omega_2') 
+ (m_q + \omega_1')(m_q + \omega_2) \nonumber \\
   &        & + (m_q + \omega_1)(m_q + \omega_2) + (m_q + \omega_1')(m_q + \omega_2')]]
\end{eqnarray}
\par

	Some features of this ``master" expression may be noted. There is  
a `natural factorization'in the variables $q_{12}$  and $q_{12}'$, except 
for the photon propagator $k^{-2}$, $(k = q_{12} - q_{12}')$.  Further, 
the two blobs are connected together by the `spectator'variable $p_3$ which                                                                                               
is on the mass shell due to the presence of $\Gamma_3^*\Gamma_3'$ in eq.(4.3). 
\par
                                               
	The time-like (pole) integrations over each of $d\sigma_{12}$ and 
$\sigma_{12}'$ can be carried out {\it exactly} a la (2.28) and its 
derivatives, since the 3D vertex function $D_{12}\phi$  in $\Gamma_3$ does 
not involve $\sigma_{12}$, etc. After this step  ${\hat q}_{12}$, 
${\hat q}_{12}'$ and ${\hat p}_3$  are the `right' 3D variables for the
`triple integration' whose essential logic may be stated as follows. The 
main strategy is to decouple the ${\hat q}$ and ${\hat q}'$ variables                               
from the photon propagator k  through the following device [19]: 
\par

	Since $k$ is basically space- like, it is a good approximation to       
replace $k^{-2}$ by ${\hat k}^{-2}$  which equals $({\hat q}_{12} - 
{\hat q}_{12}')^2$, and drop  the angular correlation in the two ${\hat q}$- 
momenta (since the error in this neglect is zero in the first order [28]). 
Next we use the inequality [28] 
\begin{equation}
(a^2 +b^2)^{-1} \leq (2ab)^{-1}; \quad a  \rightarrow |{\hat q}_{12}|, etc
\end{equation}
which ensures the necessary factorizability in the q-variables.In principle 
the corrections to this inequality can be calculated since the neglected term 
is  approximately equal to $-(a-b)^2/(4a^2b^2)$  which is still factorizable, 
but this refinement is unnecessary in view of the smallness of the e.m. effect 
itself. After this simplification the rest of the integration procedure is 
straightforward  since the ${\hat q}$ and ${\hat q}'$ integrations can be 
done analytically, and only a 1D integration over $|p_3|$ remains for 
numerical evaluation. The necessary expressions are collected in  Appendix A 
and the numerical results for all contributions are given as under.
\par
	The key parameters are the quark mass $m_q$  and the size parameter 
$\beta^2$, the latter being determined dynamically through the chain of 
eqs.(2.11-14). As noted in Sec.1 already, the mass $m_q$  which is usually 
called the `constituent' mass, should be viewed as the sum of the (flavour 
independent) `mass function' m(p) for small p, plus a small ``current mass" 
$m_c$, in the spirit of Politzer additivity [22]. The mass function m(p) was 
generated  in this BSE-cum-SDE framework through a  Dynamical Chiral Symmetry 
Breaking mechanism in a non perturbative fashion [23]. Also from some related 
quark-loop calculations with $q{\bar q}$ mesons in recent times [24,28], it 
was found that for such `low energy' processes the mass function m(p) is 
rather well approximated by m(0), so that [22], $m_q = m(0)+m_c$. Therefore 
the $d-u$ mass difference is the {\it same} at the `constituent' or at the 
`current' levels, and this is what has been denoted by $\delta m$ in the text 
(figs.2). Its smallness compared to $m_q$  justifies its neglect in all the 
functions except where it appears explicitly, viz., fig.2. We take its value 
at $\delta m = 4 MeV$, as in related calculations [24,28], while the other 
quantities are predetermined from $q{\bar q}$ [8] and $qqq$ [9] spectroscopy:
\begin{equation}
m_q = 265 MeV; \quad \beta^2 (N) = 0.052 GeV^2
\end{equation}
so that there are {\it {no free}} parameters in the entire calculation.
The results from Secs.2-4 are now summarized for (n-p) as :
\begin{equation}
N_B^{-2} = 5.5209 \times 10^{-4} GeV^{-10}; \quad [e_i = 1/3]
\end{equation}  
\begin{equation}
\delta M_{st} = +1.7134 MeV ; \quad \delta M^{\gamma} = -0.4396 MeV.
\end{equation}   
Hence 
\begin{equation}
\delta M(net) = + 1.28 MeV; \quad (vs. 1.29 MeV : Expt)
\end{equation} 
which is the principal result of this investigation.
\section {Summary and Conclusion}
\par
	This calculation fills up an important gap in the two-tier BSE
formalism under 3D kernel support (CIA) for a simultaneous investigation
of spectra and transition amplitudes of both $q{\bar q}$ and $qqq$ varieties
within a single unified framework [12,13]. To recapitulate the main points, 
the (first stage) 3D reductions of both the 2-body and 3-body BSE's had 
yielded good agreement with the respective spectra [8,9], with a common set 
of parameters $C_0 = 0.27$  and $\omega_0 = 158 MeV$ characterizing the 
structure of the non-perturbative gluon propagator, since a third parameter, 
the `constituent' mass $m_q$ needed for spectroscopy [8,9], is essentially 
the dynamical mass function $m(p)$ in the low momentum limit [22, 23]. 
\par
	More substantial tests of the formalism are expected from the (second 
stage) reconstruction of the 4D hadron-quark vertex function which carries the 
non-perturbative off-shell information in a {\it closed} form. This exercise 
was initially confined to the meson-$q{\bar q}$ vertex function whose exact 
reconstruction [12] had led to several useful checks, from 4D loop integrals
for hadronic and e.m. transition amplitudes [11,12], to like integrals probing
the momentum dependence of the quark mass function $m(p)$ which is the 
`chiral' limit ($M_\pi = 0$) [17,21,23] of the pion-quark vertex function. 
Indeed $m(p)$ acts as the form factor for loop integrals determining the 
vacuum to vacuum transitions, and is found to predict correctly several 
condensates, from the basic $<q{\bar q}>$ [23] to `induced' condensates [39], 
{\it {under one roof}}. Further tests of the hadron-quark vertex function have
come from SU(2) breaking effects like $\rho-\omega$ mixing [24] and mass 
splittings in pseudoscalar mesons [28], with only one additional parameter 
representing the $d-u$ mass difference; see also [27].
\par
	The last link in our two-tier formalism has been a reconstruction 
of the 4D baryon-$qqq$ vertex function on the lines of the 2-body case [12], 
to facilitate the evaluation of corresponding 3-body loop integrals. This
quantity was conjectured some time ago [13], but has only recently found a
rigorous derivation [29] in terms of Green's function techniques, whose main 
steps are sketched for completeness in Appendix B. The present work represents
a first application of this quantity, choosing as an example the problem of 
the $n-p$ mass difference. The physics of this problem is two-fold: i) the 
$qqq$ vertex function which is entirely determined by the {\it same} dynamics 
as for the $q{\bar q}$ case; ii) the  strong and e.m. SU(2) breaking, 
(figs.2 and 3), considered on identical lines to the corresponding problem of 
pseudoscalar SU(2) mass splittings [28]. As such we have refrained here from
giving fuller references on the $d-u$ mass difference, most of which 
originated from Weinberg's famous paper [40],  but several more references 
on the physics of the problem may be found in [28]. On the other hand the 
entire derivation shows that no free parameters are involved, so that the
final figure (4.13), although a single number, must {\it not} be treated as
an isolated quantity, but as an integral part of a much bigger package.  
\par
	As a point of detail, we should also mention our neglect of the 
condensate contributions inserted in the internal quark lines as in figs.2, 
in view of a recent finding [28] that such contributions are small within the 
BSE framework. This may not be too surprising since, unlike in the QCD-SR 
method [26] where such condensate contributions are the principal source of 
non-perturbative effects, this is no longer the case in the present BSE 
treatment which was primarily designed to incorporate non-perturbative
effects in the zeroeth order. (As a result, the condensate effects in the 
present non-perturbative scenario may well be residual).In this respect
the good agreement of our net estimate of $\delta M$, eq.(4.13), with 
experiment (without free parameters) suggests a good support for the BS 
dynamics when viewed together with other related phenomena [8,9,12,23,24,39] 
within the same dynamical framework [12,13,23,29].
\par 
	We also recapitilate the `QCD' status of this BSE formalism [12]
viv-a-vis contemporary methods like QCD sum rules [26] or chiral perturbation
theory [27]. As already explained in Sec.1, the gluon exchange character 
of the pairwise $q{\bar q}$ or $qq$ interactions lends them a natural chiral
invariance property at the input Lagrangian level with `current' quarks. In
particular, the `constituent' mass is {\it not} an input, but emerges as the
low momentum limit of the dynamical mass function $m(p)$ that characterizes
the quark propagators appearing in the 2- and 3- body BSE's, as a result
of $DB{\chi}S$ [21, 17, 19, 23], since the `current' masses of $u,d$ quarks 
give only a small additive contribution [22]. The empirical aspect of the
gluon propagator concerns only its {\it {non-perturbative}} regime which  
often requires separate parametrization even in orthodox formulations [18].
In the present formulation, its explicit parametrization with two constants 
$C_0$ and $\omega_0$) [23] is the price for a `closed form' representation 
of non-perturbative effects in the derived hadron-quark vertex function, 
but the returns are rich, e.g., the description of many items that depend 
sensitively on the details of such form factors in the quark-loop integrals
[12,23,24,28], with their structures firmly rooted in spectroscopy [8,9].
In contrast, the chiral perturbation theory [27] has a more explicit QCD 
content, but with greater emphasis on a perturbative treatment, as revealed 
by expansions in powers of small momenta and ``current" masses $m_c$ [27] 
for a systemic derivation of the low energy structure of the Green's function 
in QCD [27]. It is a powerful method, highly successful in predicting items 
like ground state masses {\it {as well as}} their splittings, but its lack of
a closed form representation prevents an equally successful prediction of 
`soft' QCD effects in enough details, such as the momentum dependence of the 
mass function, or of hadron-quark vertex functions in general, with other
observable consequences such as failure to predict the L-excited spectra [10]. 
The method of QCD sum rules [26] also shares similar features such as lack of 
closed form representation for the form factors, and while it does predict 
items like the $n-p$ mass difference, the highly indirect nature of such 
derivations [41] brings out the parametric uncertainties involved in the 
simulation of soft QCD effects.      
\par

	Finally we should like to comment on the principal motivation for 
this investigation, namely to demonstrate the practical feasibility of such 
realistic quark-loop calculations for the relativistic 3-quark problem with 
a full-fledged (BS) dynamical framework whose basic parameters are linked all 
the way to spectroscopy. The present calculation indeed suggests that not 
only quark-loops involving mesons [12,23,24,28] but even those involving the 
(less trivial) $qqq$ baryon are amenable to a similar degree of dynamical 
sophistication without excessive efforts, so that it makes sense to speak of 
an effective ``4-fermion coupling" for both $q{\bar q}$ and $qq$ pairs within 
a common parametric framework. This is somewhat remiscient of Bethe's 
``second principle" theory, originally suggested at the two-nucleon level of 
nuclear forces, now reinterpreted at the quark level, with a simple extension 
to include the antiquark in the dynamical description. (This extension would 
not make sense at the $NN$ level, since the $NN$ and $N{\bar N}$ forces are 
very different from each other). Indeed such a dynamics had been strongly 
suggested (with concrete examples) in a perspective review not too long ago 
[42], but it seemed to have gone largely by default, as evidenced by a strong 
tendency in the contemporary literature to continue to rely on ``ad-hoc form 
factors" [43] to simulate the vertex functions, instead of generating them
dynamically. Hopefully, some efforts in this direction have been recently in 
evidence [44], using the Nambu Jonalasino model [21] of contact 4-fermion
interactions. It is to be hoped that Bethe's ``second principle" perspective 
will be upheld by such investigations, at least until such time as a fully 
satisfactory solution to QCD is  forthcoming.   
\par
	
	The initial draft of this paper was prepared at the National 
Institute of Advanced Studies. We are grateful to Dr.Raja Ramanna for the 
warm NIAS hospitality. We also acknowledge Ms Chandana's help with some
`difficult' figures.

\section*{Appendix A: Evaluation of the Integral (4.5)} 
\setcounter{equation}{0}
\renewcommand{\theequation}{A.\arabic{equation}}
\par
	
	The master expression (4.8) after being substituted in the full e.m.
 self energy contribution (4.5) is integrated over each $d\sigma_{12}$ and 
 $d{\sigma_{12}}'$. The final result is 
\begin{eqnarray}
\delta M^{\gamma} & = & \sum_{123} {{2e^2} \over {9}} \tau_3 \int 
{(m_q + \omega_3)\over {2\omega_3}} {d^3{\hat p}_3\over {4\pi}} {d^3{\hat q}_{12}\over {4\pi}} 
{d^3{{\hat q}_{12}}'\over {4\pi}} {1 \over {2{\hat q}_{12} 2{{\hat q}_{12}}'}} \times \nonumber \\
 		  &   & F({\hat q}_{12}, {{\hat q}_{12}}', {\hat p}_3 )
exp{(-{2\over 3}[{{\hat q}_{12}}^2 + {{\hat q}_{12}}^{'2} + 3{{\hat p}_3}^2]/{\beta}^2)}
\end{eqnarray} 						 
where
\begin{eqnarray}
\lefteqn{F({\hat q}_{12}, {{\hat q}_{12}}', {\hat p}_3 ) = }  \nonumber \\
  & & (m_q + \omega_{12})(m_q + {\omega_{12}}') k^2 + (M\omega_3 - {1\over 2}
(M^2 - m_q^2))[(m_q + \omega_{12})^2 \nonumber \\
  & & + (m_q + {\omega_{12}}')^2 + (m_q + \omega_{12})(m_q + {\omega_{12}}') 
- (m_q + \omega_{12})^2 {D_{12}}'/{2{\omega_{12}}'} \nonumber \\
  & & - (m_q + {\omega_{12}}')^2 D_{12}/{2\omega_{12}} 
- (m_q + \omega_{12})(m_q + {\omega_{12}}')[D_{12}/{2\omega_{12}} \nonumber \\
  & & + {D_{12}}'/{2{\omega_{12}}'}] k^2 [(m_q + 2 \omega_{12})(m_q 
+ 2{\omega_{12}}') - m_q^2]/2 \nonumber \\
  & & - [(m_q + 2 \omega_{12})(m_q + 2{\omega_{12}}') - m_q^2] 
[D_{12}/{2\omega_{12}} + {D_{12}}'/{2{\omega_{12}}'}]/2  \nonumber \\
  & & -{1\over 8} {D_{12} {D_{12}}'\over {\omega_{12} {\omega_{12}}'}} + k^4
-  k^2 [D_{12}/{\omega_{12}} + {D_{12}}'/{\omega_{12}}']
\end{eqnarray} 
Using eq.(4.9), the integration over ${\hat q}_{12}$  and ${{\hat q}_{12}}'$ 
can be done independently of each other, and thus can be written in a compact 
notation as follows 
\begin{equation}
\delta M^{\gamma} = \sum_{123} (+{2\over 9}e^2 \tau_3) \int {\hat p}_3^2
d{\hat p}_3 {(m_q + \omega_3)\over {2\omega_3}} F_1 \ e^{[- {\hat p}_3^2/\beta^2]}
\end{equation}
where
\begin{eqnarray}
F_1 & = & J_{11}J_{11} + [M\omega_3 -{1\over 2}M^2 +{1\over 2}m_q^2] 
(2J_{20}J_{00} + 2J_{10}J_{10}) - 2J_{20}I_{00} \nonumber \\
    &   & - 2J_{10}I_{10} + {J_{11}}'{J_{11}}'/2 -J_{01}J_{01}m_q^2/2 
- {I_{10}}'{J_{10}}' \nonumber \\
    &   & + m_q^2 I_{00}J_{00} - I_{00}I_{00}/2 -J_{02}J_{02}/4 + I_{01}I_{01}
\end{eqnarray}       
and with $(n = 0,1,2 ; m = 0,1,2)$,
\begin{equation}
J_{nm};I_{nm} = 2^{-1/2} \int {\hat q}_{12} d{\hat q}_{12} \ e^{(-{2\over 3}
{{\hat q}_{12}}^2/\beta^2)}[{\sqrt 2}{\hat q}_{12}]^m (m_q + \omega_{12})^n
[1 ;  {1\over 2}D_{12}/{\omega_{12}}];
\end{equation}

\begin{equation}
{J_{nm}}';{I_{nm}}' = 2^{-1/2} \int {\hat q}_{12} d{\hat q}_{12} \ e^{(-{2\over 3}
{{\hat q}_{12}}^2/\beta^2)} [{\sqrt 2}{\hat q}_{12}]^m (m_q + 2 \omega_{12})^n
[1 ;  {1\over 2}D_{12}/{\omega_{12}}];
\end{equation}

\section*{Appendix B: Derivation of the qqq Vertex Structure, Eq.(2.5)} 
\subsection*{B.1: Method of Green's Functions} 
\setcounter{equation}{0}
\renewcommand{\theequation}{B.1.\arabic{equation}}
\par

	We outline here some essential steps leading to a formal derivation of 
eq.(2.5) which was written down in a semi-intuitive fashion in [13]. To that
end we shall employ the method of Green's functions for 2- and 3- particle
scattering {\it {near the bound state pole}}, since the inhomogeneous terms
are not relevant for our purposes. For simplicity we shall consider identical
{\it spinless} bosons, with pairwise BS kernels under CIA conditions [12],
first for the 2-body case for calibration, and then for the 3-body system.
\subsection*{B.2: Two-Quark Green's Function}
\setcounter{equation}{0}
\renewcommand{\theequation}{B.2.\arabic{equation}}
\par

	Apart from some results already goven in the text, we shall use the 
notation and phase conventions of [12,13] for the various quantities (momenta,
propagators, etc). The 4D $qq$ Green's function $G(p_1p_2 ; {p_1}'{p_2}')$ 
near a {\it bound} state satisfies a 4D BSE without the inhomogeneous term, 
viz. [12,13],
\begin{equation}
i(2\pi)^4 G(p_1 p_2;{p_1}'{p_2}') = {\Delta_1}^{-1} {\Delta_2}^{-1} \int
d{p_1}'' d{p_2}'' K(p_1 p_2;{p_1}''{p_2}'') G({p_1}''{p_2}'';{p_1}'{p_2}')    
\end{equation}
 where
\begin{equation}
\Delta_1 = {p_1}^2 + {m_q}^2 , 
\end{equation}
and $m_q$ is the mass of each quark. Now using the relative 4- momentum 
$q = (p_1-p_2)/2$ and total 4-momentum $P = p_1 + p_2$ 
(similarly for the other sets), and removing a $\delta$-function
for overall 4-momentum conservation, from each of the $G$- and $K$- 
functions, eq.(B.2.1) reduces to the simpler form    
\begin{equation}
i(2\pi)^4 G(q.q') = {\Delta_1}^{-1} {\Delta_2}^{-1}  \int d{\hat q}'' 
Md{\sigma}'' K({\hat q},{\hat q''}) G(q'',q')
\end{equation}
where ${\hat q}_{\mu} = q_{\mu} - {\sigma} P_{\mu}$, with 
$\sigma = (q.P)/P^2$, is effectively 3D in content (being orthogonal to
$P_{\mu}$). Here we have incorporated the ansatz of a 3D support for the
kernel $K$ (independent of $\sigma$ and ${\sigma}'$), and broken up the 
4D measure $dq''$ arising from (2.1) into the product 
$d{\hat q}''Md{\sigma}''$ of a 3D and a 1D measure respectively. We have 
also suppressed the 4-momentum $P_{\mu}$ label, with $(P^2 = -M^2)$, in 
the notation for $G(q.q')$.
\par

	Now define the fully 3D Green's function ${\hat G}({\hat q},{\hat q}')$
as [29] 
\begin{equation}
{\hat G}({\hat q},{\hat q}') = \int \int M^2 d{\sigma}d{\sigma}'G(q,q')
\end{equation}
and two (hybrid) 3D-4D Green's functions ${\tilde G}({\hat q},q')$,
${\tilde G}(q,{\hat q}')$ as
\begin{equation}
{\tilde G}({\hat q},q') = \int Md{\sigma} G(q,q');
{\tilde G}(q,{\hat q}') = \int Md{\sigma}' G(q,q');
\end{equation} 
Next, use (B.2.5) in (B.2.3) to give    
\begin{equation}
i(2\pi)^4 {\tilde G}(q,{\hat q}') = {\Delta_1}^{-1} {\Delta_2}^{-1} 
\int dq'' K({\hat q},{\hat q}''){\tilde G}(q'',{\hat q}')  
\end{equation}
Now integrate both sides of (B.2.3) w.r.t. $Md{\sigma}$ and use the result [12]
\begin{equation}
\int Md{\sigma}{\Delta_1}^{-1} {\Delta_2}^{-1} = 2{\pi}i D^{-1}({\hat q});
\quad D({\hat q}) = 4{\hat \omega}({\hat \omega}^2 - M^2/4);\quad 
{\hat \omega}^2 = {m_q}^2 + {\hat q}^2 
\end{equation}
to give a 3D BSE w.r.t. the variable ${\hat q}$, while keeping the other 
variable $q'$ in a 4D form:
\begin{equation}  
(2\pi)^3 {\tilde G}({\hat q},q') = D^{-1} \int d{\hat q}''  
K({\hat q},{\hat q}'') {\tilde G}({\hat q}'',q')
\end{equation}
Now a comparison of (B.2.3) with (B.2.8) gives the desired connection between 
the full 4D $G$-function and the hybrid ${\tilde G({\hat q}, q')}$-function: 
\begin{equation}  
2{\pi}i G(q,q') = D({\hat q}){\Delta_1}^{-1}{\Delta_2}^{-1}
{\tilde G}({\hat q},q')
\end{equation}
Again, the symmetry of the left hand side of (B.2.9) w.r.t. $q$ and $q'$ 
allows us to write the right hand side with the roles of $q$ and $q'$ 
interchanged. This gives the dual form   
\begin{equation}  
2{\pi}i G(q,q') = D({\hat q}'){{\Delta_1}'}^{-1}{{\Delta_2}'}^{-1}
{\tilde G}(q,{\hat q}')
\end{equation}
which on integrating both sides w.r.t. $M d{\sigma}$ gives
\begin{equation}  
2{\pi}i{\tilde G}({\hat q},q') = D({\hat q}'){{\Delta_1}'}^{-1}
{{\Delta_2}'}^{-1}{\hat G}({\hat q},{\hat q}'). 
\end{equation}
Substitution of (B.2.11) in (B.2.9) then gives the symmetrical form
\begin{equation}  
(2{\pi}i)^2 G(q,q') = D({\hat q}){\Delta_1}^{-1}{\Delta_2}^{-1}
{\hat G}({\hat q},{\hat q}')D({\hat q}'){{\Delta_1}'}^{-1}
{{\Delta_2}'}^{-1}
\end{equation}
Finally, integrating both sides of (B.2.8) w.r.t. $M d{\sigma}'$, we 
obtain a fully reduced 3D BSE for the 3D Green's function:
\begin{equation}  
(2\pi)^3 {\hat G}({\hat q},{\hat q}') = D^{-1}({\hat q} \int d{\hat q}''
K({\hat q},{\hat q}'') {\hat G}({\hat q}'',{\hat q}')
\end{equation}
Eq.(B.2.12) which is valid near the bound state pole (since the 
inhomogeneous term has been dropped for simplicity) expresses the desired 
connection between the 3D and 4D forms of the Green's functions; and 
eq(B.2.13) is the determining equation for the 3D form. A spectral analysis 
can now be made for either of the 3D or 4D Green's functions in the 
standard manner, viz., 
\begin{equation}  
G(q,q') = \sum_n {\Phi}_n(q;P){\Phi}_n^*(q';P)/(P^2 + M^2) 
\end{equation}
where $\Phi$ is the 4D BS wave function. A similar expansion holds for 
the 3D $G$-function ${\hat G}$ in terms of ${\phi}_n({\hat q})$. Substituting
these expansions in (B.2.12), one immediately sees the connection between 
the 3D and 4D wave functions in the form:
\begin{equation}  
2{\pi}i{\Phi}(q,P) = {\Delta_1}^{-1}{\Delta_2}^{-1}D(\hat q){\phi}(\hat q)
\end{equation}
whence the BS vertex function becomes $\Gamma = D \times \phi/(2{\pi}i)$
as found in [12]. We shall make free use of these results, taken as $qq$ 
subsystems, for our study of the $qqq$ $G$-functions in Sections 3 and 4.  

\subsection*{B.3: 3D Reduction of the BSE for 3-Quark G-function}
\setcounter{equation}{0}
\renewcommand{\theequation}{B.3.\arabic{equation}}
\par

	As in the two-body case, and in an obvious notation for various 
4-momenta (without the Greek suffixes), we consider the most general 
Green's function $G(p_1 p_2 p_3;{p_1}' {p_2}' {p_3}')$ for 3-quark 
scattering {\it near the bound state pole} (for simplicity) which allows       
us to drop the various inhomogeneous terms from the beginning. Again we 
take out an overall delta function $\delta(p_1 + p_2 + p_3 - P)$ from the
$G$-function  and work with {\it two} internal 4-momenta for each of the 
initial and final states defined as follows [13]:
\begin{equation}  
{\sqrt 3}{\xi}_3 =p_1 - p_2 \ ; \quad  3{\eta}_3 = - 2p_3 + p_1 +p_2
\end{equation}
\begin{equation}  
P = p_1 + p_2 + p_3 = {p_1}' + {p_2}' + {p_3}'
\end{equation}
and two other sets ${\xi}_1,{\eta}_1$ and ${\xi}_2,{\eta}_2$ defined by 
cyclic permutations from (B.3.1). Further, as we shall consider pairwise
kernels with 3D support, we define the effectively 3D momenta ${\hat p}_i$, 
as well as the three (cyclic) sets of internal momenta 
${\hat \xi}_i,{\hat \eta}_i$, (i = 1,2,3) by [13]:
\begin{equation}
{\hat p}_i = p_i - {\nu}_i P \ ;\quad  {\hat {\xi}}_i = {\xi}_i - s_i P\  ;
\quad
{\hat {\eta}}_i - t_i P 
\end{equation}
\begin{equation}  
{\nu}_i = (P.p_i)/P^2\  ;\quad s_i = (P.\xi_i)/P^2 \ ;\quad t_i = 
(P.\eta_i)/P^2 \end{equation}
\begin{equation}  
{\sqrt 3} s_3 = \nu_1 - \nu_2 \ ;\quad 3 t_3 = -2 \nu_3 + \nu_1 + \nu_2 \ 
\ ( + {\rm cyclic permutations})
\end{equation}

The space-like momenta ${\hat p}_i$ and the time-like ones $\nu_i$ 
satisfy [13] 
\begin{equation}  
{\hat p}_1 + {\hat p}_2 + {\hat p}_3 = 0\  ;\quad \nu_1 + \nu_2 + \nu_3 = 1
\end{equation}
Strictly speaking, in the spirit of covariant instantaneity, we should 
have taken the relative 3D momenta ${\hat \xi},{\hat \eta}$ to be in the 
instantaneous frames of the concerned pairs, i.e., w.r.t. the rest frames
of $P_{ij} = p_i +p_j$; however the difference between the rest frames of 
$P$ and $P_{ij}$  is small and calculable [13], while the use of a common 
3-body rest frame $(P = 0)$ lends considerable simplicity and elegance to 
the formalism.   
\par

	We may now use the foregoing considerations to write down the BSE 
for the 6-point Green's function in terms of relative momenta, on closely 
parallel lines to the 2-body case. To that end note that the 2-body 
relative momenta are $q_{ij} = (p_i - p_j)/2 = {\sqrt 3}{\xi_k}/2$, where 
(ijk) are cyclic permutations of (123). Then for the reduced $qqq$ Green's
function, when the {\it last} interaction was in the (ij) pair, we may use 
the notation $G(\xi_k \eta_k ; {\xi_k}' {\eta_k}')$, together with `hat' 
notations on these 4-momenta when the corresponding time-like components 
are integrated out. Further, since the pair $\xi_k,\eta_k$ is 
{\it {permutation invariant}} as a whole, we may choose to drop the index 
notation from the complete $G$-function to emphasize this symmetry as and 
when needed. The $G$-function for the $qqq$ system satisfies, in the 
neighbourhood of the bound state pole, the following (homogeneous) 4D BSE
for pairwise $qq$ kernels with 3D support:
\begin{equation}  
i(2\pi)^4 G(\xi \eta ;{\xi}' {\eta}') = \sum_{123}
{\Delta_1}^{-1} {\Delta_2}^{-1} \int d{{\hat q}_{12}}'' M d{\sigma_{12}}''
K({\hat q}_{12}, {{\hat q}_{12}}'') G({\xi_3}'' {\eta_3}'';{\xi_3}' {\eta_3}')
\end{equation}
where we have employed a mixed notation ($q_{12}$ versus $\xi_3$) to stress
the two-body nature of the interaction with one spectator at a time, in a 
normalization directly comparable with eq.(B.2.3) for the corresponding 
two-body problem. Note also the connections 
\begin{equation}  
\sigma_{12} = {\sqrt 3}{s_3}/2   ;\quad 
{\hat q}_{12} = {\sqrt 3}{{\hat \xi}_3}/2  ; \quad {\hat \eta}_3 = 
-{\hat p}_3, \quad etc 
\end{equation}  
The next task is to reduce the 4D BSE (B.3.7) to a fully 3D form through a 
sequence of integrations w.r.t. the time-like momenta $s_i,t_i$ applied 
to the different terms on the right hand side, {\it {provided both}} 
variables are simultaneously permuted. We now define the following fully 
3D as well as mixed (hybrid) 3D-4D $G$-functions according as one or more 
of the time-like $\xi,\eta$ variables are integrated out:
\begin{equation}  
{\hat G}({\hat \xi} {\hat \eta};{\hat \xi}' {\hat \eta}') = 
\int \int \int \int ds dt ds' dt' G(\xi \eta ; {\xi}' {\eta}')  
\end{equation}  
which is $S_3$-symmetric.
\begin{equation}  
{\tilde G}_{3\eta}(\xi {\hat \eta};{\xi}' {\hat \eta}') = 
\int \int dt_3 d{t_3}' G(\xi \eta ; {\xi}' {\eta}');
\end{equation}  
\begin{equation}  
{\tilde G}_{3\xi}({\hat \xi}  \eta;{\hat \xi}' {\eta}') = 
\int \int ds_3 d{s_3}' G(\xi \eta ; {\xi}' {\eta}');
\end{equation} 
The last two equations are however {\it not} symmetric w.r.t. the 
permutation group $S_3$, since both the variables ${\xi,\eta}$ are not 
simultaneously transformed; this fact has been indicated in eqs.(B.3.10-11) 
by the suffix ``3" on the corresponding (hybrid) ${\tilde G}$-functions,
to emphasize that the `asymmetry' is w.r.t. the index ``3". We shall term 
such quantities ``$S_3$-indexed", to distinguish them from $S_3$-symmetric 
quantities as in eq.(B.3.9). The full 3D BSE for the ${\hat G}$-function is 
obtained by integrating out both sides of (B.3.7) w.r.t. the $st$-pair 
variables $ds_i d{s_j}' dt_i d{t_j}'$ (giving rise to an $S_3$-symmetric 
quantity), and using (B.3.9) together with (B.3.8) as follows:
\begin{equation}  
(2\pi)^3 {\hat G}({\hat \xi} {\hat \eta} ;{\hat \xi}' {\hat \eta}') = 
\sum_{123} D^{-1}({\hat q}_{12}) \int d{{\hat q}_{12}}'' 
K({\hat q}_{12}, {{\hat q}_{12}}'') {\hat G}({\hat \xi}'' {\hat \eta}'';
{\hat \xi}' {\hat \eta}')  
\end{equation}   
This integral equation for ${\hat G}$ which is the 3-body counterpart of
(B.2.13) for a $qq$ system in the neighbourhood of the bound state pole, 
is the desired 3D BSE for the $qqq$ system in a {\it {fully connected}}
form, i.e., free from delta functions. Now using a spectral decomposition 
for ${\hat G}$ 
\begin{equation}   
{\hat G}({\hat \xi} {\hat \eta};{\hat \xi}' {\hat \eta}')
= \sum_n {\phi}_n( {\hat \xi} {\hat \eta} ;P)
{\phi}_n^*({\hat \xi}' {\hat \eta}';P)/(P^2 + M^2)
\end{equation}   
on both sides of (B.3.12) and equating the residues near a given pole
$P^2 = -M^2$, gives the desired equation for the 3D wave function $\phi$ 
for the bound state in the connected form:
\begin{equation}   
(2\pi)^3 \phi({\hat \xi} {\hat \eta} ;P) = \sum_{123} D^{-1}({\hat q}_{12})
\int d{{\hat q}_{12}}'' K({\hat q}_{12}, {{\hat q}_{12}}'')
\phi({\hat \xi}'' {\hat \eta}'' ;P)
\end{equation}   
Now the $S_3$-symmetry of $\phi$ in the $({\hat \xi}_i, {\hat \eta}_i)$ pair 
is a very useful result for both the solution of (B.3.14) {\it and} for the 
reconstruction of the 4D BS wave function in terms of the 3D wave function 
(B.3.14), as is done in the subsection below.
\subsection*{B.4: Reconstruction of the 4D BS Wave Function}
\setcounter{equation}{0}
\renewcommand{\theequation}{B.4.\arabic{equation}}
\par

	We now attempt to {\it re-express} the 4D $G$-function given by 
(B.3.7) in terms of the 3D ${\hat G}$-function given by (B.3.12), as the 
$qqq$ counterpart of the $qq$ results (B.2.12-13). To that end we adapt 
the result (B.2.12) to the hybrid Green's function  of the (12) subsystem 
given by ${\tilde G}_{3 \eta}$, eq.(B.3.10), in which the 3-momenta 
${\hat \eta}_3,{{\hat \eta}_3}'$ play a parametric role reflecting the 
spectator status of quark $\# 3$, while the {\it active} roles are played 
by $q_{12}, {q_{12}}' = {\sqrt 3}(\xi_3,{\xi_3}')/2$, for which the analysis 
of subsec.B.2 applies directly. This gives 
\begin{equation} 
(2{\pi}i)^2 {\tilde G}_{3 \eta}(\xi_3 {\hat \eta}_3; 
{\xi_3}' {{\hat \eta}_3}') 
= D({\hat q}_{12}){\Delta_1}^{-1}{\Delta_2}^{-1}
{\hat G}({\hat \xi_3} {\hat \eta_3}; {\hat \xi_3}' {\hat \eta_3}')
D({{\hat q}_{12}}'){{\Delta_1}'}^{-1}{{\Delta_2}'}^{-1}
\end{equation}
where on the right hand side, the `hatted' $G$-function has full 
$S_3$-symmetry, although (for purposes of book-keeping) we have not 
shown this fact explicitly by deleting the suffix `3' from its 
arguments. A second relation of this kind may be obtained from (B.3.7)
by noting that the 3 terms on its right hand side may be expressed in 
terms of the hybrid ${\tilde G}_{3 \xi}$ functions vide their definitions 
(B.3.11), together with the 2-body interconnection between $(\xi_3,{\xi_3}')$ 
and $({\hat \xi}_3,{{\hat \xi}_3}')$ expressed once again via (B.4.1), but 
without the `hats' on $\eta_3$ and ${\eta_3}'$. This gives
\begin{eqnarray}
({\sqrt 3} \pi i)^2 G(\xi_3 \eta_3; {\xi_3}'{\eta_3}')
&=& ({\sqrt 3} \pi i)^2 G(\xi \eta; {\xi}'{\eta}')\nonumber\\
&=& \sum_{123} {\Delta_1}^{-1}{\Delta_2}^{-1} (\pi i {\sqrt 3})
\int d{{\hat q}_{12}}'' M d{\sigma_{12}}''
K({\hat q}_{12}, {{\hat q}_{12}}'') 
G({\xi_3}'' {\eta_3}'';{\xi_3}' {\eta_3}')\nonumber\\   
&=& \sum_{123} D({\hat q}_{12}) {\Delta_1}^{-1}{\Delta_2}^{-1}
{\tilde G}_{3 \xi}({\hat \xi}_3  \eta_3; {{\hat \xi}_3}' {{\eta}_3}')
{{\Delta_1}'}^{-1} {{\Delta_2}'}^{-1}  
\end{eqnarray}
where the second form exploits the symmetry between $\xi,\eta$ and 
$\xi',\eta'$. 
\par

	At this stage, unlike the 2-body case, the reconstruction of the
4D Green's function is {\it {not yet}} complete for the 3-body case, as 
eq.(B.4.2) clearly shows. This is due to the {\it truncation} of Hilbert 
space implied in the ansatz of 3D support to the pairwise BSE kernel $K$ 
which, while facilitating a 4D to 3D BSE reduction without extra charge, 
does {\it not} have the {\it complete} information to permit the {\it reverse}
transition (3D to 4D) without additional assumptions; see [29] for details. 
The physical reasons for the 3D ansatz for the BSE kernel have been discussed 
in detail elsewhere [23,29], vis-a-vis contemporary approaches. Here we look 
upon this ``inverse" problem as a purely {\it mathematical} one.  
\par

	We must now look for a suitable  ansatz for the quantity 
${\tilde G}_{3 \xi}$ on the right hand side of (B.4.2) in terms of {\it known}
quantities, so that the reconstructed 4D $G$-function satisfies the 3D 
equation (B.3.12) exactly, as a ``check-point" for the entire exercise. We 
therefore seek a structure of the form 
\begin{equation}
{\tilde G}_{3 \xi}({\hat \xi}_3  {\eta}_3; {{\hat \xi}_3}' {{\eta}_3}')
= {\hat G}({{\hat \xi}_3} {\hat \eta}_3; {{\hat \xi}_3}' {{\hat \eta}_3}')
\times F(p_3, {p_3}')    
\end{equation}
where the unknown function $F$ must involve only the momentum of the 
spectator quark $\# 3$. A part of the $\eta_3, {\eta_3}'$ dependence has 
been absorbed in the ${\hat G}$ function on the right, so as to satisfy 
the requirements of $S_3$-symmetry for this 3D quantity [29]. 
\par

	As to the remaining factor $F$, it is necessary to choose its
form in a careful manner so as to conform to the conservation of 
4-momentum for the {\it free} propagation of the spectator between two
neighbouring vertices, consistently with the symmetry between $p_3$ 
and ${p_3}'$. A possible choice consistent with these conditions is
the form (see [29] for details):
\begin{equation}
F(p_3, {p_3}') = C_3 {\Delta_3}^{-1} {\delta}(\nu_3 - {\nu_3}') 
\end{equation}
Here ${\Delta_3}^{-1}$ represents the ``free" propagation of quark $\# 3$ 
between successive vertices, while $C_3$ represents some residual effects 
which may at most depend on the 3-momentum ${\hat p}_3$, but must satisfy 
the main constraint that the 3D BSE, (B.3.12), be {\it explicitly} satisfied.
\par

	To check the self-consistency of the ansatz (B.4.4), integrate
both sides of (B.4.2) w.r.t. $ds_3 d{s_3}' dt_3 d{t_3}'$ to recover the 
3D $S_3$-invariant ${\hat G}$-function on the left hand side. Next, in 
the first form on the right hand side, integrate w.r.t. $ds_3 d{s_3}'$ 
on the $G$-function which alone involves these variables. This yields
the quantity ${\tilde G}_{3 \xi}$. At this stage, employ the ansatz 
(B.4.4) to integrate over $dt_3 d{t_3}'$. Consistency with the 3D BSE, 
eq.(B.3.12), now demands 
\begin{equation}
C_3 \int \int d\nu_3 d{\nu_3}' {\Delta_3}^{-1} \delta(\nu_3 - {\nu_3}')
= 1 ; (since dt = d\nu) 
\end{equation}
The 1D integration w.r.t. $d\nu_3$ may be evaluated as a contour 
integral over the propagator ${\Delta}^{-1}$ , which gives the pole 
at $\nu_3 = {\hat \omega}_3/M$, (see below for its definition). Evaluating 
the residue then gives 
\begin{equation}
C_3 = i \pi / (M {\hat \omega}_3 ) ;  \quad
{{\hat \omega}_3}^2 = {m_q}^2 + {{\hat p}_3}^2
\end{equation}
which will reproduce the 3D BSE, eq.(B.3.12), {\it exactly}! Substitution
of (B.4.4) in the second form of (B.4.2) finally gives the desired 3-body 
generalization of (B.2.12) in the form 
\begin{equation}
3 G(\xi \eta; \xi' \eta') = \sum_{123} D({\hat q}_{12}) \Delta_{1F} 
\Delta_{2F} D({{\hat q}_{12}}') {\Delta_{1F}}' {\Delta_{2F}}' 
{\hat G}({\hat \xi_3} {\hat \eta_3}; {\hat \xi_3}' {\hat \eta_3}')
[\Delta_{3F} / (M \pi {\hat \omega}_3)]      
\end{equation}
where for each index, $\Delta_F = - i {\Delta}^{-1}$ is the 
Feynman propagator.
\par

	To find the effect of the ansatz (B.4.4) on the 4D BS 
{\it {wave function}} $\Phi(\xi \eta; P)$, we do a spectral reduction 
like (B.3.13) for the 4D Green's function $G$ on the left hand side of 
(B.4.2). Equating the residues on both sides gives the desired 4D-3D 
connection between $\Phi$ and $\phi$:
\begin{equation}
\Phi(\xi \eta; P) = \sum_{123} D({\hat q}_{12}){\Delta_1}^{-1}{\Delta_2}^{-1}
\phi ({\hat \xi} {\hat \eta}; P) \times 
\sqrt{{\delta(\nu_3 -{\hat \omega}_3/M)} \over{M {\hat \omega}_3 
{\Delta}_3}} 
\end{equation}
From (B.4.8) and eq.(2.1) of the text, we infer the structure of the 
baryon-$qqq$ vertex function $V_3$ as given in eq.(2.5) of the text. For
a detailed discussion of the significance of this result, vis-a-vis 
contemporary approaches, see [29].

\section*{Figure Captions}
\par
Fig.1: Diagrams for BS normalization of Baryon-$qqq$ vertex function. 1(a)
shows quark $\# 1$ emitting a zero momentum photon ($k=0$); its last $qq$
interaction was with $\# 2$, while $\# 3$ is the spectator. 1(b) is the 
same diagram with the roles of $\# 1$ and $\# 2$ interchanged. 1(c) 
denotes schematically two more such pairs of diagrams obtained with cyclical
permutations of the indices (123) in pairs. The 4-momenta on the quark lines
are shown as used in the text.
\par
Fig.2: Diagrams for the two-point interactions of the quark lines with the
mass shift operator $-\delta m {\tau_3}^{(1)}/2$ in place of the photon 
in fig.1, but otherwise with identical topological correspondence of 
figs.2(a,b,c) to figs.1(a,b,c).
\par
Fig.3: Diagrams for the e.m. self-energy of the $uud$ (proton) configuration. 	
3(III) is shown in detail with full momentum markings as employed in the 
text, and corresponds to quark $\# 3$ as the spectator, while the quark lines
$\# 1$ and $\# 2$ are joined by a transverse photon line. Similarly 3(I) 
and 3(II) correspond to $\# 1$ and $\# 2$ respectively as spectators in turn. 
Note that, unlike in fig.1 and fig.2, the interchange of $\# 1$ and $\# 2$
in fig.3(III) does not give a new configuration. 

\par

{\bf Note} : 
Due to lack of adequate software for a proper `DVI' rendering of the three 
figures above, it has not been found possible to include the same in this
file. The inconvenience on this account is regretted. However, any interested 
reader will find it quite easy to reconstruct the three ``quark loop" figures
(in each of which there are {\it three} internal quark lines between two 
baryon vertex blobs) on the basis of the information supplied in the captions
above. As a further guidance, the interested reader may refer to ref. [28] 
which contains similar figures with two quark lines each between the 
corresponding vertex blobs of the pseudoscalar mesons concerned.

\end{document}